\documentclass{revtex4}
\usepackage{graphicx,caption,subcaption,float}

\graphicspath{ {./images/} }

\begin{document}

\title{Search for crucial events in physiological processes}
\author{Yawer H. Shah ${^1}$, Paolo Grigolini ${^1 }$}

\address{$^1$ Center for Nonlinear Science, University of North Texas, P.O. Box 311427,
	Denton, Texas 76203-1427, USA\\ }
\maketitle
\section*{Abstract}
The main purpose of this paper is to attract the attention of researchers working in the field of  physiological processes, towards crucial events. Crucial events are often confused with
extreme events thereby generating the misleading impression that their treatment should be based on quantum mechanical formalism. We show that crucial events are invisible and should not be confused with catastrophes.  Crucial events are generated by self-organization processes yielding a form of swarm intelligence, and signal their action 
with fluctuations characterized by anomalous scaling and $1/f$ spectrum. The existence or the lack of crucial events can be revealed with an entropic method of analysis called the Diffusion Entropy Analysis (DEA). However, anomalous scaling and $1/f $ spectrum are not a compelling signature of efficient self-organization, and physiological processes with anomalous scaling and $1/f$ noise spectrum without crucial events are a signature of collapsing physiological organizations.  In the case of physiological processes like cancer dynamics, the existence of crucial events is a signal of intelligence that must be destroyed rather than reinforced.

\section{Introduction}
This short review is devoted to discuss a still open problem of great importance: to establish with a special method of analysis whether  a given physiological process is affected  by disease-inducing perturbations.We illustrate the main ideas of the method focusing on a physiological signal of the cardiological processes. The field of research on extreme events is attracting an increasing number of researchers. The idea of extreme events is illustrated in several books \cite{sornette1,sornette2}, in publications \cite{havlin}, and in review papers \cite{ginestra}, \cite{earth}. Extreme events may be the results of cooperation between different units of the same system, thereby being a manifestation of self organization \cite{sornette2}. However extreme events are visible being identified with catastrophes, such as earthquakes and  stock market crashes\cite{sornette1}.  As clearly stressed in the review paper of \cite{earth} the term ``extreme event" is commonly adopted to denote unanticipated natural events of perceived high impact. 

The definition of extreme events as unpredictable phenomena involves the use of the concept of probability connecting these events with the field of complex networks. Ref. \cite{havlin} is a remarkably interesting example of this direction of research. The tools of probability theory are used adopting quantum mechanics and the quantum mechanical transport of information based on the entanglement of quantum bits. This research direction is becoming extremely popular and the very recent review paper of \cite{ginestra} is a remarkably interesting example of this research direction, using quantum mechanical formalism as a fundamental tool. Entanglement, originally proposed by Einstein, Podolsky and Rosen \cite{epr} to stress that Quantum Mechanics is an incomplete theory, is now adopted to explain the surprising role of information in the field of Complex Networks.

One important question that we have for the advocates of extreme events is how they can be used to account for the complexity of ECGs and EEGs, which can be easily represented as one-dimensional time series. The main goal of this short review is to prove that the physiological processes are characterized by invisible events, called by us, ``crucial events". These crucial events are often invisible and are not catastrophes. In fact, pathology maybe characterized by the lack of these invisible crucial events. 

To understand the importance of crucial events we should introduce our readership to the important property of scaling. The intuitive definition of scaling can be given imagining a runner 
in the one-dimensional space $x$. Let us imagine that the runner moves from the origin $x = 0$ at time $t = 0$  with velocity $v$. At time $t$ we have $x = vt$. Let us imagine that the runner makes frequent changes of direction. In this case we have \cite{wiener}, $x \propto \sqrt{t}$. Thus we can unify these two different condition with the equation

\begin{equation} \label{walker}
x \propto t^{\delta},
\end{equation}
with $\delta = 1$ when the random walker never changes direction and $\delta = 0.5$, when the random walker changes walking direction in a completely random way. This work is devoted to discuss a condition intermediate between the perfect order, $\delta = 1$, and the complete disorder, $\delta = 0.5$. 

The outline of the paper is as follows. Section (\ref{extremevents}) is devoted to emphasizing the difference between crucial events that are not yet properly taken into account in the physiological literature and the extreme events that are, on the contrary,  attracting an exponentially increasing attention. Section (\ref{diffusionentropy}) is devoted to illustrate the important tool of Diffusion Entropy Analysis (DEA) that we use to detect crucial events. We devote Section (\ref{fbm})
to discuss a phenomenon of anomalous scaling characterized by the lack of crucial events and we show how to use DEA to prove this important property. Finally we devote Section (\ref{end}) to concluding remarks.

\section{Extreme events and crucial events} \label{extremevents}
To afford an intuitive illustration of crucial events, we rest on the fundamental work of Pomeau and Manneville \cite{pm}. A Rayleigh-B\'enard convection is the coordinated movement of a fluid trapped between two thermally conducting plates. The fluid is heated from below. As a result of this heating process when the temperature gradient r is sufficiently large a coordinated motion is generated under the form of either clockwise or counterclockwise regular rotations. When the control parameter $r$ exceeds a critical value $r_T$, the regular oscillations are replaced by short bursts of disorder (turbulence). These short bursts of disorder are our crucial events.  To facilitate the creation of crucial events, rather than using the popular Manneville map, adopted for instance by Zumofen and Klafter \cite{zumofenklafter} we follow the prescription of \cite{idealized} called by us idealized version of the Manneville map. We consider the equation of motion

\begin{equation}
\dot y = \lambda y^z,
\end{equation}
where $y$ is a variable moving in  the interval $I = [0,1]$. The choice of the initial condition is totally random, with the points of this interval sharing the same probability. The particle moves from this initial condition towards $y= 1$. The time spent by the particle to reach the border is called $\tau$. The process is totally deterministic. Randomness is introduced in this picture by selecting a new random initial condition. With easy analytical calculations we find that the distribution density $\psi(\tau)$ gets the inverse power law structure:

\begin{equation} \label{waiting}
\psi(\tau) = (\mu -1) \frac{T^{\mu-1}}{(\tau + T)^{\mu}}, 
\end{equation}

where

\begin{equation}
T  \equiv \frac{\mu -1}{\lambda}
\end{equation}

and

\begin{equation}
\mu   \equiv \frac{z}{z-1}. 
\end{equation}

The value of the complexity index $\mu$ ranges from $\mu = 1$ to $\mu = \infty$. However the values $\mu= 2$ and $\mu =3$ have a special meaning, connected to the moments 

\begin{equation}
m_n \equiv <\tau^n>.
\end{equation}
If $ \mu < 2$ only the zero order moment exists, as a consequence of the fact that the waiting time distribution is normalized. If $2 < \mu < 3$, the first moment is
given by
\begin{equation}
< \tau > = \frac{T}{\mu -2}.
\end{equation} 
The condition $3 < \mu < 4$ is characterized by a finite value for both the first and the second moment.  For this reason, the condition $\mu = 3$  is frequently considered to be the border between complex and ordinary statistical physics. This is not fully correct as discussed in Ref. \cite{annunziato}.  A time series $\{\xi(t)\}$ hosting crucial events does not generate a
stationary correlation function $C(t_1, t_2) = <\xi(t_1) \xi)t_2)>$, namely correlation functions depending only on  $|t_1 - t_2|$.

There exist two special values of $\mu$, $\mu = 2$ and $\mu = 3$.
The value $\mu = 2$ is a transition from a regime with $\mu < 2$, characterized by perennial aging \cite{mirko}, namely, a regime where the correlation function $C$ never becomes stationary  to a regime with $2 < \mu < 3$, where the correlation function becomes stationary in the long-time limit. Physiological processes are frequently found to work in the regime
$2 < \mu < 3 $ \cite{synchronization}. Callum et al. \cite{callum} discuss geophysical tremors as dynamical processes occurring in the regime $2 < \mu < 3$, thereby sharing the same properties as physiological signals, a condition that may be violated by the extreme events \cite{sornette1}. 

It is important to stress that the waiting time distribution density of Eq. (\ref{waiting}) can be derived directly from the renewal theory of Cox \cite{cox}, as illustrated in \cite{mirko}. 
In his book Cox introduces his readership to crucial events interpreted by him as failure events. A machine created at time $t= 0$ may fail at time $t > 0$. The age dependent failure rate 
$g(t)$ is defined by Cox as
\begin{equation}
g(t) = \frac{\psi(t)}{\Psi(t)}, 
\end{equation}
where $\psi(t)$ is the time distance between birth, $t = 0$, and failure time at $t> 0$.

The function $\Psi(t)$ is the corresponding survival probability, namely the probability that no failure has occurred up to a time $t$,
\begin{equation}
\Psi(t) = \int_{t}^{\infty} dt' \psi(t').
\end{equation}

This makes it possible for us to write $g(t)$ as

\begin{equation}
g(t) = - \frac{\dot \Psi(t)}{\Psi(t)},
\end{equation}

yielding

\begin{equation} \label{intuitive}
\Psi(t) = e^{-\int_{0}^{t} g(t') dt'}. 
\end{equation}

The exponential condition $\Psi(t) = e^{-r_0 t}$ is easily derived by assuming that $g(t)$ is constant. Research work fitting the goals of insurance companies seems to be oriented to make the conjecture that $g(t)$ is an exponentially increasing function of time. The research work of our group, on the contrary, focuses on the assumption 
\begin{equation} \label{genialassumption}
g(t) = \frac{r_0}{1 + r_1t}.
\end{equation}

With very simple algebra, we find that Eq. (\ref{genialassumption}) yields the same waiting time distribution density as the idealized Manneville map, Eq. (\ref{waiting}) with 
\begin{equation}
T \equiv 1/r_1
\end{equation} 
and
\begin{equation}
\mu \equiv 1 + \frac{r_0}{r_1}.
\end{equation}

The assumption that $g(t)$ decreases rather than increasing in time is certainly questionable if it is referred to human beings, interpreted as machines designed by engineers. It becomes 
reasonable and attractive when it is applied to physiological processes. In that case we can interpret the failure as a biological malfunction, instantaneously repaired, a condition that seems to be compatible with the opinion of biologists \cite{instantaneousrepair} that the neuron to neuron interaction may also involve the instantaneous malfunction repair. 
In this case a crucial event would be the signature of healthy dynamics, in a striking contrast with the interpretation of a crucial events as a catastrophe. 

What about the quantum mechanical approach \cite{havlin,ginestra}  to critical phenomena? We believe in the remarks made by Francis Heylighen \cite{francis}. This author discusses
the Rayleigh-B\'enard convection phenomenon that we are using to explain the occurrence of crucial events emphasizing the important role of wave function collapses, claiming that
it is a mysterious property of quantum mechanics that may be connected to his approach to complexity. We are rephrasing his remarks stressing that crucial events may be a manifestation of wave function collapses.

\section{Diffusion Entropy} \label{diffusionentropy}
To shed light on crucial events, we now describe our proposal to detect them, despite being invisible, a property establishing a striking difference between crucial events and catastrophes. The tool that we propose to detect crucial events is the Diffusion Entropy Analysis (DEA) \cite{DEA}. The method is based on the observation of a time series $\xi(n)$,
the interbeat time between the beat n and the beat n+1. The inter-beat times have a mean value close to 1 second. We plot the difference between the inter-beat time and the mean value, generating the fluctuation process illustrated in Fig. 1. We divide the vertical axis into stripes of size s, and we record the time at which the signal enters a given stripe. We focus our attention on the time spent by the signal within a given stripe. The time at which the signal enters a given stripe (or gets out from it)  is an event.   

\begin{figure}[h]
    \centering
    \includegraphics[width=0.4\textwidth]{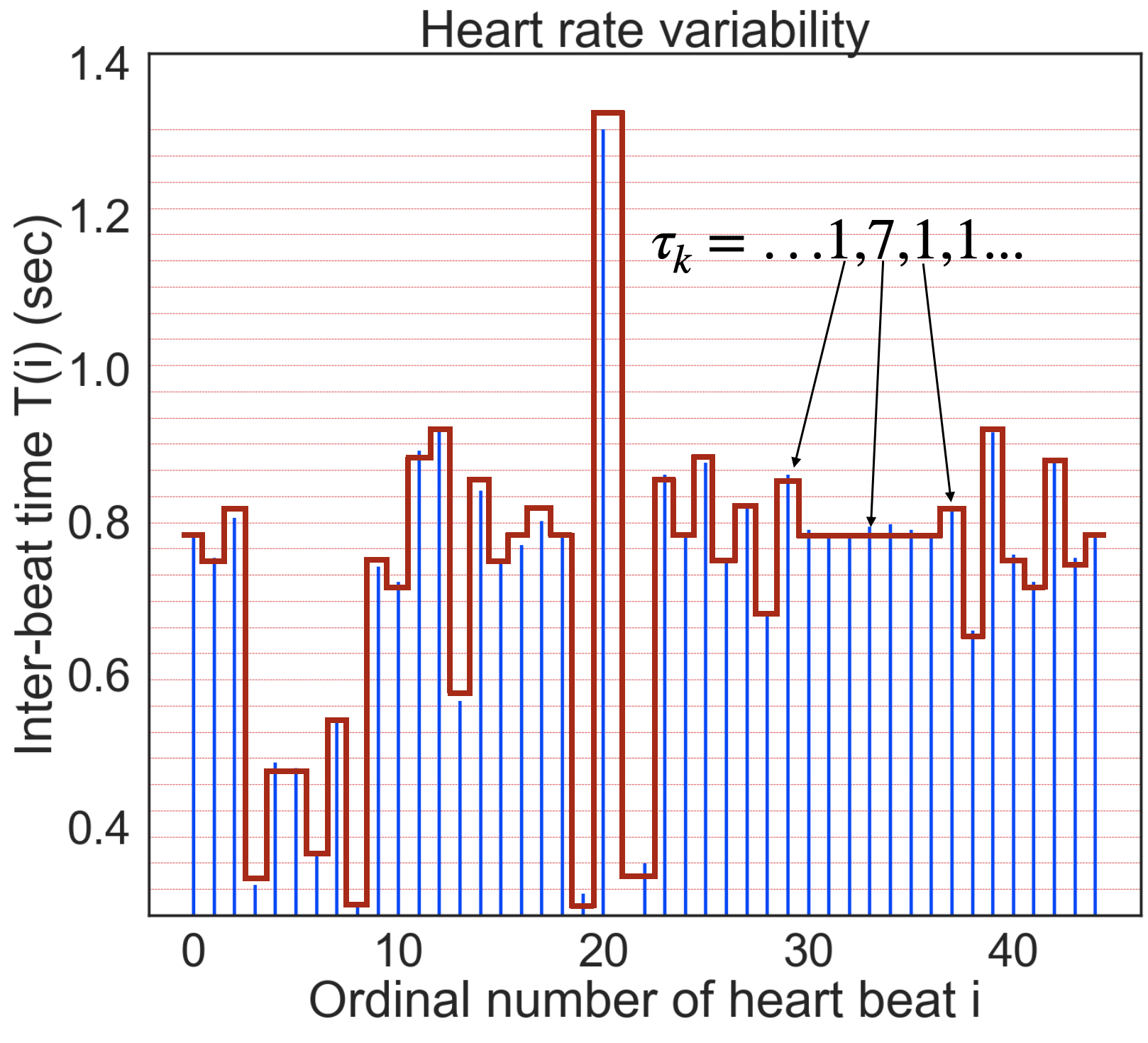}
    \caption{The sequence $\tau_k$ is extracted from the inter-beat times by recording the time the signal spends in a given stripe. The stripe numbers are marked on the right. The stripe width s = 0.033 s.}
    \label{fig1}
\end{figure}

The time distance between the entrance and the exit time is called $\tau$. If the entrance and exit times are crucial, the corresponding waiting time distribution density is expected to 
be given by Eq. (\ref{waiting}). with $1  < \mu < 3$. The reason for this condition is afforded by \cite{giacomo}, which shows that for $\mu \to 3$ the scaling $\delta \to 0.5$. 
However, we cannot rule out the possibility that also values of $\mu$ larger than $3$ may generate values of the scaling $\delta$ slightly larger than $0.5$ as a consequence of the fact that the transition from $\mu < 3$ to $\mu > 3$ is perceived at values of  $\mu$ slightly larger than $3$ \cite{annunziato} as discussed in detail in \cite{callum}. 

We have to stress an important issue concerning the size of the stripes that allows us to find the correct scaling. The size of the stripes is a fraction of the span of intensity of the fluctuations. As a consequence some events generated by the crossings of the stripes are not crucial. The time distance between two consecutive non-crucial events is given by an exponential waiting time distribution density and their contribution to scaling is 
\begin{equation}
\delta = \frac{1}{2}.
\end{equation}
This is a well known property \cite{schlesinger} that can be derived from Eq. (\ref{intuitive}) by  setting $g(t) = r_0$, namely by assuming that $r_1 = 0$ in Eq. (\ref{genialassumption}).

This leads us to reiterate that the crucial events are invisible, namely that the experimental signal illustrated in Fig. 1, denoted by the symbol $\xi$ can be written as follows:

\begin{equation}  \label{experimental} 
\xi(t) = p_1 \xi_{NC}(t) + p_2 \xi_{C}(t). 
\end{equation}
The symbol $\xi_{(NC}(t)$denotes the contribution to the physiological time series generated by non-crucial events, either Poisson or of infinite memory. As we see in $\ref{fbm}$, the signals with a perennially stationary correlation function do not host crucial events.

We assume that the physiological process under study generates a scaling lager than $\delta = 0.5$.  In the long-time limit the random walk $x(t)$ obtained by the integration over the experimental time series of Eq.  (\ref{experimental}) 
\begin{equation}
x(t) = \int_{0}^{t} dt' \xi(t')
\end{equation}
will depart from the origin according to the larger scaling of crucial events. In other words, a representation more accurate of the procedure described in this paper is obtained by replacing Eq. (\ref{walker}) with

\begin{equation}
x(t) \propto   p_1 t^{\delta_{NC}} + p_2 t^{\delta_{C}}. 
\end{equation}

Note that
\begin{equation}
p_1 + p_2 = 1
\end{equation}
and that  the analysis of real cardiological data done in \cite{DEA}  shows that the transition from the healthy to the pathological condition is not signaled only by a small reduction of the scaling $\delta$. It is also signaled by a reduction of $p_2$ and corresponding increase of $p_1$. 
With this paper we want to attract the attention of the readership with main interest on physiological processes, to the fact that this transition to a pathological condition is not only due to a slight increase of $\delta_{C}$ but also due to an increase of the concentration of non-crucial events, including infinite memory as well as Poisson events. 

Note that the research work done by our group on physiological processes led us to the conclusion that $2 < \mu <3$ for both ECGs \cite{DEA} and EEGs \cite{ecg}. In both cases

\begin{equation} \label{anomalous} 
\delta = \frac{1}{\mu -1}, 
\end{equation}
which for $2 < \mu <3$ makes $\delta_{C} > \delta_{NC} = 0.5$. This explains why the method of DEA \cite{DEA} makes it possible to detect the invisible crucial events, which are characterized by  a scaling larger than non-crucial events. However, it is to be noted that the method of DEA proves that $\delta_{C}$ $>$ $\delta_{NC}$ explaining why the pathological signals are more renewal than the healthy signals.

\section{Anomalous Scaling without Crucial Events} \label{fbm}

It is important to stress that the anomalous scaling of Eq. (\ref{anomalous}) is not limited to ECGs and  EEGs. It is a sign of ``intelligence" of complex systems ranging from the brain to social systems, generated by the cooperative interaction of their units, \cite{collectiveintelligence},   \cite{pingitore}, with $2 < \mu <3$ with $\mu$ very close to $2$.  This is in line with the statement that the brain is a source of maximal intelligence, yielding $\delta = 1$. Recent work affords support to the conjecture that also the motility of cancer cells share the condition  of intelligence corresponding to $2 < \mu < 3$. The work of \cite{chiara} based on the observation of motility of cancer cells is leading to $2 < \mu <3$, with the preliminary result of $\mu = 2.5$. The motility of cancer cells is an issue of increasing interest, as shown by the recent paper of \cite{motility}.  Of course, the use of the statistical analysis proposed by this paper should contribute to the search of a method to kill the intelligence of cancer cells.

To explain this challenging issue, we have to notice that the work of \cite{maurizio,herbert}  shows that the seed germinating in an environment not affording the proper nutrients and the heart of an healthy patient under homeopathic neuropathy of increasing severity makes a transition from a condition hosting crucial events to an anomalous scaling without crucial events.  To properly interpreting the nature of this transition we review the theoretical work of our group on the origin of the $1/f$ noise. 

The $1/f$ spectrum generated by crucial events  has the form \cite{mirko}

\begin{equation}
S(f) = \frac{1}{f^{\beta}}
\end{equation}
with 
\begin{equation}
\beta = 3 - \mu,
\end{equation}
thereby generating $1/f$ noise when $\mu =2$.

As discussed in \cite{maurizio} the FBM $1/f$ noise is characterized by
\begin{equation}
\beta = 2 H -1,
\end{equation}
thereby generating the ideal $1/f$ noise when $H= 1$. 

In the case of physiological processes observed through ECGs and EEGs the transition from anomalous scaling generated by crucial events to anomalous scaling without crucial events is a sign of a pathology that may be cured with noninvasive stimuli inducing crucial events \cite{pingitore}. 
\begin{figure}[ht]
    \centering
    \includegraphics[width=0.3\textwidth]{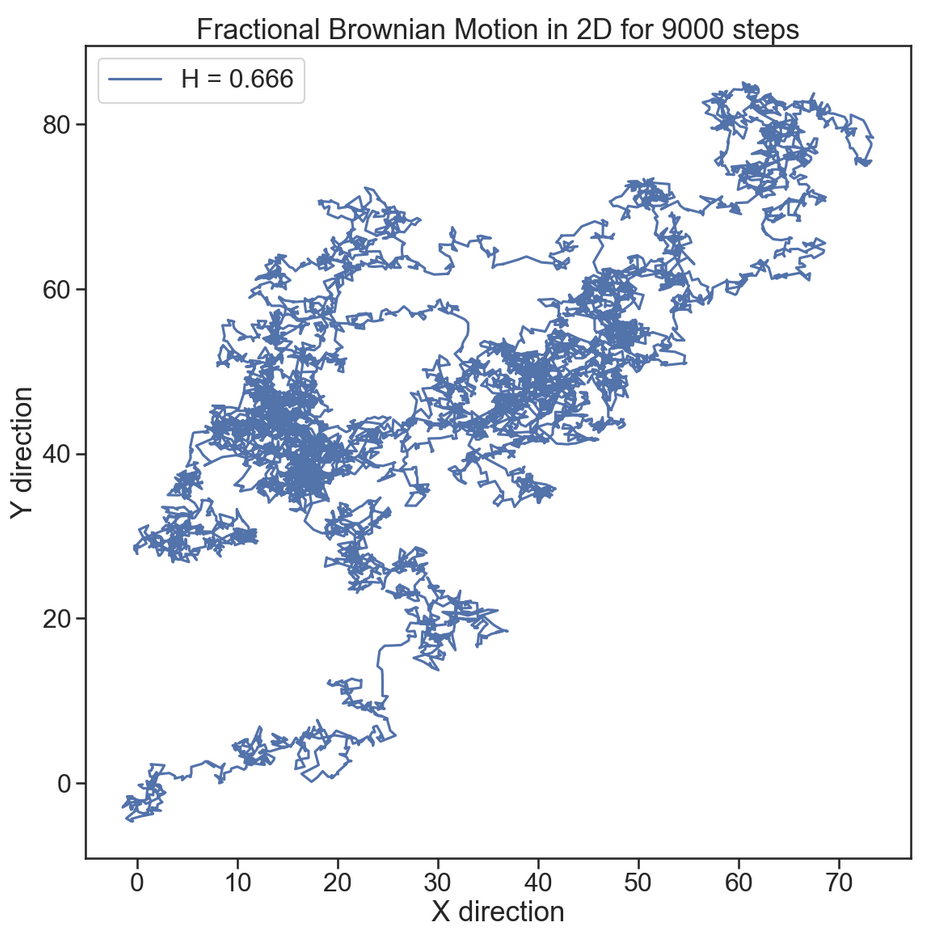}
    \caption{Fig 2: Shown here is the trajectory of a particle moving freely in 2 dimensions, generated by projecting Fractional Brownian Motion along the X and Y directions.}
    \label{fig2}
\end{figure}

The theoretical benefit of the observation of the motility of cancer cells with the method of analysis of this paper is that the pathological effects of a transition from a condition dominated by crucial events to an anomalous scaling without crucial events \cite{maurizio,herbert} may suggest the discovery of a technique to kill the crucial events of cancer cells.
Fig, 2 illustrates the motility of a cell with anomalous scaling $ H = 0.666$. Actually this is a motility without crucial events, generated from Fractional Brownian Motion (FBM). 
This form of infinite memory was proposed by Mandelbrot and Weiss \cite{mandelbrot}. Our group \cite{krockin} emphasized the quantum mechanical origin of this form of anomalous scaling and the work of \cite{garlandfbm} showed that the adoption of DEA \cite{DEA} with the stripes illustrated in Section \ref{diffusionentropy} has the effect of killing this form of anomalous diffusion yielding $\delta = 0.5$, proving the lack of intelligence of this form of anomalous diffusion. 

\begin{figure}[ht]
    \centering
    \includegraphics[width=0.3\textwidth]{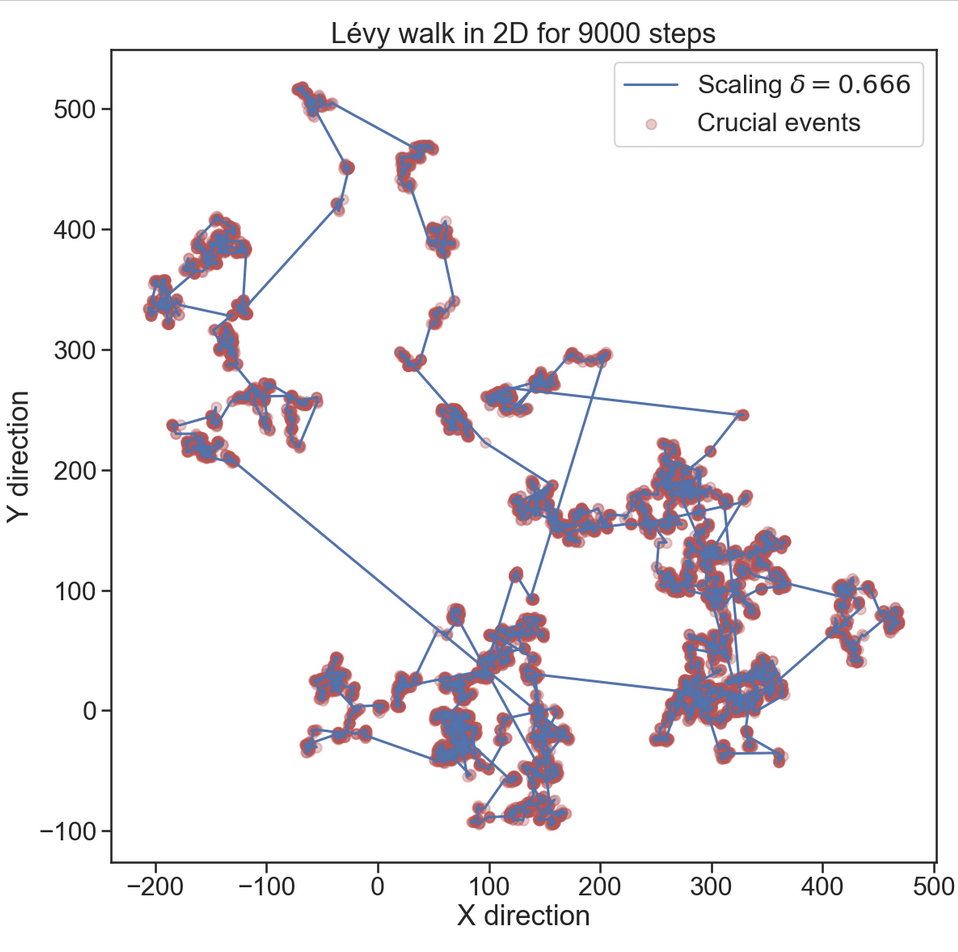}
    \caption{The diffusion generated by a 2-dimensional Lévy walker with an inverse power law index $\mu = 2.5$, simulating the movement of a cancer cell in a 2-dimensional fluid medium.}
    \label{fig3}
\end{figure}

FBM is a process with infinite memory generated using quantum mechanical arguments \cite{garlandfbm}. It can be derived from the adoption of quantum mechanical 
arguments \cite{krockin}.  Thus according the theoretical proposal of this paper quantum mechanics does not generate complexity, in conflict with the advocates of quantum mechanics \cite{havlin}, \cite{ginestra}, but an apparent form of complexity, that may be interpreted as a kind of organization collapse. 

Fig. 3 was generated to mimic the dynamics of cancerous cell \cite{chiara} assuming that they behave as a L\'evy walk, which is a two-dimensional diffusion with crucial events. In the case of Fig. 3, $\mu = 2.5$. Eq. (\ref{anomalous}) assigns to this trajectory the same scaling as that of Fig. 2, $\delta = H = 0.666$. The cell moves with constant velocity and from time  to time it changes flying direction. It is remarkable that the changes of direction are visible crucial events, becoming invisible when we observe the x and y projections of the trajectory. DEA affords the same results, $\delta = 0.666$ for both directions as well as for the two dimensional motion. 

\section{Concluding Remarks} \label{end}
Crucial events play an important role in the case of ordinary physiological processes. The main role of bio-medicine should be to recover the missing crucial events, using non invasive stimuli, for instance music, rich of crucial events. The fight against cancer should be based on the opposite role of killing the crucial events of the intelligent malignant cells. 
It is not yet known what is the dynamical process killing crucial events. We are convinced that its discovery will lead also to an efficient method to fight cancer. Another interesting open question is that whether or not the increase of $p_1$ may lead to the increase of fbm. We note that according to Sinn \cite{sinn} the transition of Fractional Brownian Motion from one stripe to the neighboring stripe is exponential.

\textit{Acknowledgments}: We thank  ARO  and NIH for financial support through grant W911NF-23-2-0247 and sub-award GMO:240910 PO: 0000003121

\end{document}